\documentclass{sigchi-ext}

\usepackage[T1]{fontenc}
\usepackage{cite}
\usepackage{textcomp}
\usepackage[scaled=.92]{helvet} 
\usepackage{graphicx} 
\usepackage{balance}  
\usepackage{booktabs} 
\usepackage{ccicons}  
\usepackage{ragged2e} 

\def\plaintitle{On the Relationship Between Explanations, Fairness Perceptions, and Decisions}

\def\emptyauthor{}
\def\plainkeywords{Algorithmic decision-making; human-AI complementarity; explanations; fairness; perceptions; reliance}

\title{On the Relationship Between Explanations, Fairness Perceptions, and Decisions}

\numberofauthors{3}

\author{%
    \alignauthor{%
        \textbf{Jakob Schoeffer}\\
        \affaddr{Karlsruhe Institute of Technology} \\
        \affaddr{Karlsruhe, Germany} \\
        \email{jakob.schoeffer@kit.edu}
    } \vfil
    \alignauthor{%
        \textbf{Maria De-Arteaga*}\\
        \affaddr{University of Texas at Austin}\\
        \affaddr{Austin, TX, USA}\\
        \email{dearteaga@mccombs.utexas.edu}
    } \vfil
    \alignauthor{%
        \textbf{Niklas Kuehl*}\\
        \affaddr{Karlsruhe Institute of Technology}\\
        \affaddr{Karlsruhe, Germany}\\
        \email{niklas.kuehl@kit.edu}
    }
    \vfil
    \alignauthor{%
        * denotes equal contribution
    }
}

\definecolor{linkColor}{RGB}{6,125,233}
\hypersetup{%
  pdftitle={\plaintitle},
  pdfauthor={\emptyauthor},
  pdfkeywords={\plainkeywords},
  bookmarksnumbered,
  pdfstartview={FitH},
  colorlinks,
  citecolor=black,
  filecolor=black,
  linkcolor=black,
  urlcolor=linkColor,
  breaklinks=true,
}

\begin{document}

\CopyrightYear{2022}
\setcopyright{rightsretained}
\conferenceinfo{ACM CHI 2022 Workshop on Human-Centered Explainable AI (HCXAI)}{May 12--13, 2022, New Orleans, LA, USA}
\isbn{978-1-4503-6819-3/20/04}
\doi{https://doi.org/10.1145/3334480.XXXXXXX}

\copyrightinfo{\acmcopyright}

\maketitle

\RaggedRight{} 

\begin{abstract}
    It is known that recommendations of AI-based systems can be incorrect or unfair.
    Hence, it is often proposed that a human be the final decision-maker.
    Prior work has argued that explanations are an essential pathway to help human decision-makers enhance decision quality and mitigate bias, i.e., facilitate human-AI complementarity.
    For these benefits to materialize, explanations should enable humans to \emph{appropriately rely} on AI recommendations and override the algorithmic recommendation when necessary to increase distributive fairness of decisions.
    The literature, however, does not provide conclusive empirical evidence as to whether explanations enable such complementarity in practice.
    In this work, we (a) provide a conceptual framework to articulate the relationships between explanations, fairness perceptions, reliance, and distributive fairness, (b) apply it to understand (seemingly) contradictory research findings at the intersection of explanations and fairness, and (c) derive cohesive implications for the formulation of research questions and the design of experiments.
\end{abstract}


\section{Introduction}
Among many other desiderata \cite{langer2021we}, it is often assumed in the XAI literature that explanations should enable humans to assess the fairness of AI recommendations, and to ultimately make better and fairer decisions \cite{dodge2019explaining,kuppa2020black,arrieta2020explainable,das2020opportunities,gilpin2018explaining,gerlings2020reviewing,du2020fairness,ferreira2020evidence,sokol2019counterfactual,slack2020fooling}.

\paragraph{Prior findings are inconclusive}

However, as of today, there is no conclusive empirical evidence showing that explanations facilitate human-AI complementarity.
Prior work has found that explanations can influence people's fairness perceptions towards AI models and their predictions in positive or negative ways (e.g., \cite{binns2018s,dodge2019explaining,shulner2022fairness,lakkaraju2020fool}).
Other findings suggest that explanations may (e.g., \cite{chu2020visual,lai2019human}) or may not (e.g., \cite{bansal2021does,poursabzi2021manipulating,green2019principles,alufaisan2020does}) lead to enhanced human-AI performance.

\marginpar{%
  \vspace{-100pt} \fbox{%
    \begin{minipage}{0.925\marginparwidth}
      \textbf{Contributions} \\
      \vspace{1pc} \textbf{1. Proposing a framework:} We propose a framework to make explicit the relationships between explanations, fairness perceptions, reliance on AI advice, and distributive fairness. \\
      \vspace{1pc} \textbf{2. Enabling dialogue between prior works:} Our framework enables us to articulate a dialogue between (seemingly contradictory) prior works and identify research gaps. \\
      \vspace{1pc} \textbf{3. Deriving research questions:} Given explanations can mislead perceptions, we ask whether this can in turn mislead reliance on AI advice in detriment of distributive fairness.
    \end{minipage}}
    }

\paragraph{Explanations can be misleading}
Deriving conclusions from existing findings is further complicated by evidence that explanations can mislead people's beliefs \cite{lakkaraju2020fool,pruthi2019learning,chromik2019dark}, even in cases where there is no intention to manipulate \cite{ehsan2021explainability}.
Lakkaraju and Bastani \cite{lakkaraju2020fool} construct high-fidelity explanations to deceive people into trusting models that make decisions based on sensitive information (e.g., race or gender) by leveraging correlations between legitimate and sensitive features.
This way, people can be nudged into perceiving a model as procedurally fair, when in reality it is \emph{not} fair.
However, to the best of our knowledge there is a lack of research studying how such miscalibrated perceptions influence people's reliance on AI recommendations as well as potential effects on distributive fairness.

\paragraph{A holistic view is needed}

\begin{figure*}
    \centering
    \includegraphics[width=0.8\textwidth]{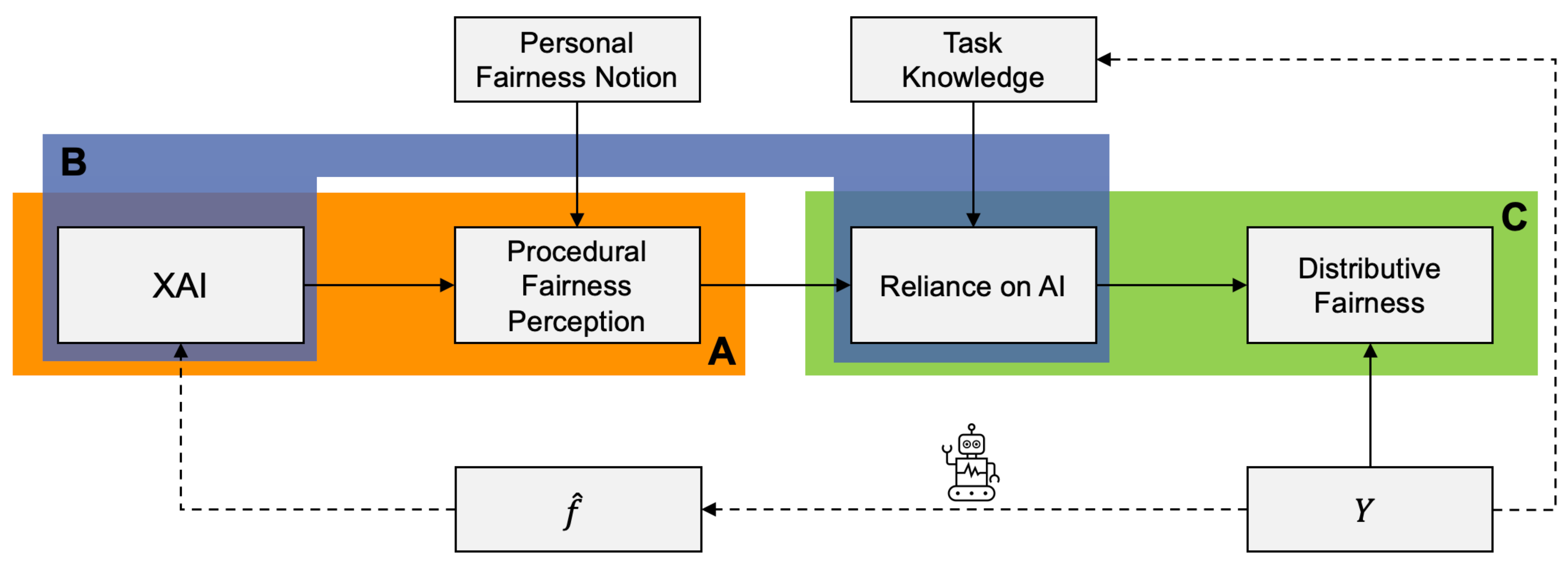}
    \caption{Conceptual framework on the interplay of explanations (XAI), procedural fairness perceptions, reliance on AI, and distributive fairness. Dashed lines indicate ``brittle'' relationships, $Y$ is the gold standard, and $\hat{f}$ is the functional representation of the AI model.}
    \label{fig:framework}
\end{figure*}

We aim to make sense of scattered findings at the intersection of explanations and fair decision-making.
Specifically, we propose a conceptual framework (see Fig.~\ref{fig:framework}) to better understand the mechanisms through which explanations may affect decisions.
To that end, we make explicit the relationships between explanations, procedural fairness perceptions, reliance on AI, and distributive fairness.
We show that the proposed framework enables us to articulate a dialogue between prior works and identify gaps that require further research. In particular, we show that prior literature has focused on different individual parts of a bigger system but that a more comprehensive lens---such as the one enabled by our proposed framework---is required to understand the effects of explanations on AI reliance and distributive fairness.

\paragraph{Pathway from explanations to distributive fairness}
Based on the application of our framework, we identify a pathway from explanations to distributive fairness, mediated by perceptions of procedural fairness and their effect on AI reliance.
We show that previous research has only studied portions of this path, and argue for its importance in the appropriate characterization of explanations' role in fair decision-making.
In particular, we conjecture that miscalibrated fairness perceptions (e.g., due to misleading explanations) may influence reliance on AI in undesirable ways, by making people adopt incorrect or override correct AI recommendations.
This lends support to the hypothesis that there is a disconnect between what explanations provide and the fairness benefits they claim.
As there is little knowledge on this interplay, we are interested in answering the following research question:

\begin{center}
\fbox{
\begin{minipage}{0.8\columnwidth}
    \textbf{RQ:} Given that explanations can mislead perceptions of fairness, (how) does this, in turn, mislead adoption/overriding behavior of AI recommendations in detriment of distributive fairness?
\end{minipage}}
\end{center}

In this workshop paper, we introduce our conceptual framework and apply it to better understand previous findings.
Addressing the above research question through a human subject study is part of our in-progress work.

\section{Conceptual Framework}

\marginpar{%
  \vspace{-220pt} \fbox{%
    \begin{minipage}{0.925\marginparwidth}
    \textbf{Definitions} \\
    \vspace{1pc}
    \textit{Procedural Fairness Perceptions:} Whether people think that the underlying AI's decision-making procedures are fair. (e.g.,~\cite{wang2020factors}) \\
    \vspace{1pc}
    \textit{Distributive Fairness:} The magnitude of disparities in error rate distributions across demographic groups. (e.g.,~\cite{chouldechova2017fair}) \\
    \end{minipage}}
    }

We propose a conceptual framework (see Fig.~\ref{fig:framework}) to make explicit the relationships between explanations \emph{(XAI)}, procedural fairness perceptions, reliance on AI recommendations, and distributive fairness.
While our framework may not  capture \emph{every} factor possibly at play, it aims to capture the primary factors considered in the literature.

\marginpar{%
  \vspace{-50pt} \fbox{%
    \begin{minipage}{0.925\marginparwidth}
    \textbf{Hypothesis} \\
    \vspace{1pc}
    Unwarranted high (low) perceptions will lead to unwarranted adherence (overrides) of AI recommendations, which make explanations an unreliable mechanism towards improving distributive fairness.
    \end{minipage}}
    }

\paragraph{Known relationships}
From prior work (e.g., \cite{grgic2018human,grgic2018beyond,grgic2020dimensions,lakkaraju2020fool,binns2018s,dodge2019explaining}), we know that explanations affect people's procedural fairness perceptions (i.e., whether people think that the underlying AI's decision-making procedures are fair).
Especially the revelation of sensitive features (e.g., gender or race) being used in the process appears to have significant effects \cite{grgic2018beyond,lakkaraju2020fool,van2019crowdsourcing}.
We further know that there are several human-specific predictors of fairness perceptions~\cite{starke2021fairness,dodge2019explaining}, which we subsume under \emph{Personal Fairness Notion}.
This may include, e.g., individuals' stance towards affirmative action~\cite{holzer2000assessing}, but may also vary across demographics~\cite{pierson2017demographics,grgic2020dimensions}.
Finally, by \emph{distributive fairness} we mean the magnitude of disparities in error rate distributions across demographic groups (e.g., males and females)~\cite{chouldechova2017fair}.

\paragraph{``Brittle'' relationships}
Our framework also includes several ``brittle'' relationships, indicated by dashed lines.
First, the relationship between $Y$ and the functional representation of the AI model $(\hat{f})$ depends on the architecture, performance, and underlying data of the employed AI model (indicated by a robot icon in Fig.~\ref{fig:framework}).
Second, the relationship between $\hat{f}$ and \emph{XAI} is ambiguous because an AI model (or its predictions) can be explained in a multitude of ways, or even independent of $\hat{f}$~\cite{eiband2019impact}---even when explanations are honest~\cite{lakkaraju2020fool,arrieta2020explainable}.
Third, task knowledge may or may not represent $Y$ (i.e., knowledge can be ``good'' or ``bad'').

\paragraph{Hypothesized relationships}
The relationship between fairness perceptions and reliance on AI (i.e., whether to adopt or override AI recommendations) is seldom touched upon in the XAI literature.
We assume a relationship to exist.
In particular, we conjecture that higher procedural fairness perceptions may be associated with increased adoption of AI recommendations---even if unwarranted.
This would be problematic insofar as perceptions can be manipulated through explanations (as discussed earlier): \emph{inappropriate reliance}~\cite{lee2004trust} might be a consequence.

\section{Applying Our Framework}
We apply our framework to previous findings, inferring that they can be divided into three groups, based on the subset of relationships that were examined (A, B, or C in Fig.~\ref{fig:framework}).

\marginpar{%
  \vspace{-130pt} \fbox{%
    \begin{minipage}{0.925\marginparwidth}
    \textbf{Takeaways} \\
    \vspace{1pc}
    \textbf{For researchers:} 
    When proposing novel XAI techniques, researchers should show how their methods do \emph{not} lead to unwarranted reliance on AI recommendations. \\
    \vspace{1pc}
    \textbf{For auditors:}
    Auditors should take a holistic perspective and assess the effects of explanations along \emph{all} dimensions of our framework (i.e., perceptions, reliance, and distributive fairness). \\
    \vspace{1pc}
    \textbf{For policy makers:}
    Knowing that explanations can be misleading, demanding a ``right to explanations'' is \emph{not} a sufficient condition for ensuring fair, accountable, and transparent use of AI in decision-making.
    \end{minipage}}
    }

\paragraph{A: XAI and fairness perceptions}
A first set of works have studied the relationship between explanations and people's perceptions.
Lakkaraju and Bastani \cite{lakkaraju2020fool}, e.g., construct explanations based on sensitive vs. relevant features and show that they can be used to mislead people into trusting untrustworthy models.
Similarly, Pruthi et al. \cite{pruthi2019learning} manipulate attention-based explanations such that people can be deceived into thinking that a model does not rely on sensitive information (e.g., gender) when in fact it does.
Binns et al. \cite{binns2018s} compared fairness perceptions across different explanation styles and scenarios---with inconclusive findings.
Dodge et al. \cite{dodge2019explaining} find that people perceive global and local explanations differently, but also conclude that the effect of explanations depends on ``the kinds of fairness issues and user profiles.''
Similarly, Shulner-Tal et al. \cite{shulner2022fairness} found that some explanations ``are more beneficial than others,'' but perceptions mainly depend on ``the outcome of the system.''

\paragraph{B: XAI and reliance on AI}
Another set of works have examined how explanations may impact people's reliance on AI \cite{schemmer2022should}.
Poursabzi-Sangdeh et al. \cite{poursabzi2021manipulating} analyzed human-AI decision-making for the case of house price estimation and found that performance did \emph{not} increase in the presence of ex\-planations---likely due to information overload.
For a similar task, however, Hemmer et al. \cite{hemmer2022effect} found that human-AI complementarity \emph{is} possible in the presence of unique human contextual information.
Green and Chen \cite{green2019principles} confirmed that explanations did not improve human performance, and Liu et al.~\cite{liu2021understanding} found that interactive explanations did not remedy this.
A similar study by Alufaisan et al. \cite{alufaisan2020does} found no conclusive evidence of explanations' influence on decision accuracy either and showed that explanations did not enable humans to detect when the AI was correct or incorrect.
Bansal et al.~\cite{bansal2021does} did observe complementarity improvements in the presence of AI augmentation, but explanations only led to over-reliance on AI advice.
On the other hand, Lai and Tan \cite{lai2019human} found that providing explanations and AI predictions can enhance human decision-making for the task of deception detection.

\paragraph{C: Reliance on AI and distributive fairness}
Several prior works have addressed the interplay of humans' reliance on AI recommendations and fairness of outcomes.
Peng et al. \cite{peng2019you} identify different types of biases in AI-based hiring decisions and find that balancing gender representations when showing potential hires to human decision-makers can correct biases in instances where humans do not exhibit persistent preferences.
Peng et al. \cite{peng2022investigations} also investigated how an AI model's predictive performance and biases may transfer to humans; one of the core findings being that different model architectures have different effects on team performance and potential mitigation of biases.
In the realm of child maltreatment screening, De-Arteaga et al. \cite{de2020case} found that call workers changed behavior in the presence of an AI recommendation, and that they were less likely to adopt incorrect AI advice.
Green and Chen \cite{green2019disparate}, however, in a different risk assessment case, found that humans under-performed the AI even when presented with its advice, were unable to evaluate both their own and the AI's performance, and biases against Black people were amplified through the use of AI recommendation.

In our in-progress work, we jointly consider relationships A, B, and C, in order to empirically examine explanations' effects on AI reliance and distributive fairness.

\balance{} 

\bibliographystyle{SIGCHI-Reference-Format}
\bibliography{bibliography}

\end{document}